\documentclass[useAMS,usenatbib]{mn2e}

 \usepackage{times}
 \usepackage{graphicx}
\newcommand{\APer}{$\alpha$ Per}

\title[The ages of L dwarfs]
{The ages of L dwarfs
\thanks{Based on observations made with the United Kingdom Infrared
Telescope, operated by the Joint Astronomy Centre on behalf of the
U.K. Particle Physics and Astronomy Research Council.}}
\author[R. F. Jameson et al.]{R. F. Jameson$^{1}$\thanks{E-mail:
rfj@star.le.ac.uk}, N. Lodieu$^{2,1}$, S. L. Casewell$^{1}$, 
N. P. Bannister$^{1}$, and P. D. Dobbie$^{3,1}$ \\
$^{1}$Department of Physics and Astronomy, University of Leicester, 
University Road, Leicester LE1 7RH, UK \\
$^{2}$Instituto de Astrof\'isica de Canarias, V\'ia L\'actea s/n,
E-38205 La Laguna, Tenerife, Spain \\
$^{3}$Anglo-Australian Observatory, P.O. Box 296, Epping 1710, Australia
\\}
\begin{document}

\date{Accepted \today{}. Received \today{}; in original form \today{}}

\pagerange{\pageref{firstpage}--\pageref{lastpage}} \pubyear{2007}

\maketitle

\label{firstpage}

\begin{abstract}
We present a new method to derive the age of young ($<$0.7 Gyr) L dwarfs based on their
near-infrared photometry, colours, and distances. The method is
based on samples of L dwarfs belonging to the Upper Sco association
(5 Myr), the Alpha Per (85 Myr) and Pleiades (125 Myr) clusters, and
the Ursa Major (400 Myr) and Hyades (625 Myr) moving groups.
We apply our method to a number of interesting objects in the 
literature, including a known L dwarf binary, L dwarf companions,
and spectroscopic members of the young $\sigma$ Orionis cluster.

\end{abstract}


\section{Introduction}

Low-mass stars and brown dwarfs undergo a significant change in luminosity
over the first hundred million years (hereafter Myr) of their life
\citep{baraffe98,burrows01}. As brown dwarfs fade inexorably with time, 
estimating their age is important to infer fundamental parameters such 
as the mass. A large number of age diagnostics and methods exist, including 
the main-sequence turn-off \citep{mermilliod81}, the lithium depletion 
boundary \citep*{rebolo92}, rotation and activity \citep{randich01},
kinematics \citep{reid02,dahn02}, and belonging to a moving group 
\citep{montes01}. However uncertainties remain both on the 
observational and theoretical sides: for example ages derived from 
the main-sequence turn-off method tend to be half those 
inferred from the lithium depletion boundary \citep{jeffries01}.

Large-scale sky surveys have uncovered a large population of ultracool
dwarfs (defined as dwarfs with spectral types later than M7)
\citep[e.g.][]{delfosse99,kirkpatrick00,geballe02}. These discoveries required 
a new class of objects cooler than M dwarfs to be defined, the L dwarfs. L dwarfs were proposed by 
\citet{kirkpatrick99} and \citet{martin99} to
describe the disappearance of titanium oxide and vanadium oxide 
and the onset of metal alkali at optical and infrared wavelengths.
Old field L dwarfs represent a mixture of very-low-mass stars and 
brown dwarfs \citep{kirkpatrick00}. They have typical effective 
temperatures between 2200 and 1400 Kelvins \citep{basri00,leggett00}
and exhibit red optical and near-infrared colours \citep{knapp04}. 
There are currently about 500 L dwarfs\footnote[1]{See
http://spider.ipac.caltech.edu/staff/davy/ARCHIVE/, a webpage
dedicated to L and T dwarfs maintained by C.\ Gelino, D.\ Kirkpatrick,
and A.\ Burgasser} known, mainly uncovered by large-scale sky surveys 
\citep[e.g.][]{kirkpatrick00,chiu06} and proper motion studies 
\citep[e.g.][]{cruz03}, including 76 with parallaxes 
\citep{perryman97,dahn02,vrba04}.

In this paper we present a scheme to infer the age of L dwarfs
using their near-infrared photometry, assuming
that they are single objects with a known distance. We have 
chosen five samples of young, intermediate-age, and old late-type 
dwarfs in clusters and moving groups to derive an age-colour
relationship independent of theoretical models.
In the first part of this paper we describe the samples used in
our study (Sect.\ \ref{age:sample}), including field L dwarf 
members of the Hyades and Ursa Major moving groups, Pleiades 
proper motion members, \APer{} photometric candidates, and 
Upper Sco spectroscopic members. In the second part
we discuss the method to determine the age of L dwarfs
(Sect.\ \ref{age:determination}). Finally, we apply our method to 
several L dwarfs with known distance published to date 
(Sect.\ \ref{age:application}).

%
%
\begin{table}
\begin{center}
  \caption{Summary of the four L dwarf members of the Hyades
moving group. Magnitudes from 2MASS and infrared spectral types are
listed for each object along with its distance derived from
parallax measurement.}
  \label{tab_age:HyadesMG_memb}
  \begin{tabular}{@{\hspace{0mm}}l c c c c c@{\hspace{0mm}}}
  \hline
2MASS J\ldots{}  &   $J$  &   $H$  & $K_{s}$ & d$_{p}$       & SpT   \\
  \hline
00361617$+$1821104  & 12.466 & 11.588 & 11.058 & 8.8 $\pm$0.1 & L4    \\
00325937$+$1410371  & 16.830 & 15.648 & 14.946 & 33.2$\pm$6.9 & L8    \\
01075242$+$0041563  & 15.824 & 14.512 & 13.709 & 15.6$\pm$1.2 & L5.5  \\
08251968$+$2115521  & 15.100 & 13.792 & 13.028 & 10.7$\pm$0.1 & L6    \\
12171110$-$0311131  & 15.860 & 15.748 & 15.887 & 11.0$\pm$0.3 & T7.5  \\
\hline
  \end{tabular}
  \end{center}
\end{table}

%
%
\section{The samples}
\label{age:sample}
%

%
%
\subsection{Field L and T dwarfs with known distances}
\label{age:field}

The full catalogue of field L and T dwarfs can be retrieved from the
DwarfArchive.org webpage\footnotemark[1].
Among the 487 known L dwarfs as of November 2006 (starting date of
this project), 76 had measured parallaxes and proper motions
\citep{perryman97,dahn02,vrba04}
but their age was, and remains for the large majority, unknown. 
We have used those parallaxes to infer the absolute $K$ magnitudes 
(M$_{K}$) of each single L dwarf (star symbols 
in Fig.\ \ref{fig_age:CMD}), i.e.\ L dwarfs not reported to date 
as belonging to multiple systems by high-resolution imaging surveys
\citep[see review by][]{burgasser06a}. The $J-K_{s}$ colours (on the 2MASS system) taken 
from the L and T dwarf archive were transformed into the Mauna Kea 
Observatory system \citep[MKO;][]{tokunaga02} using
equations detailed in  \citet{stephens04} for direct 
comparison with photometric data on open clusters extracted from 
the WFCAM Science Archive 
(Hambly et al., subm.\ to MNRAS)\footnote[2]{The archive 
can be accessed at http://surveys.roe.ac.uk/wsa/index.html}
based on the UKIRT Infrared Deep Sky Survey 
\citep{lawrence07}\footnote[3]{More details on the project can 
be found at http://www.ukidss.org/}. All $JHK$ photometry in this paper is on the MKO system unless
otherwise stated.

%
%
\subsection{The Hyades moving group}
\label{age:Hyades}

The Hyades galactic cluster is the nearest bound star cluster and has 
been extensively studied \citep*{leggett89,reid92,reid93,reid99a}. 
It has been found to be of approximate solar metallicity,  \citet{cayreldestrobel85} determine a value of [Fe/H] = $+$0.12, while \citet{perryman98} 
find 0.15\@. 
Recent age estimates range from 500 to 900 Myr \citep{barry81,kroupa95}. 
The age estimate used in this paper was determined by \citet{perryman98} who fit  theoretical isochrones which include the 
effects of convective overshoot to the {\it{Hipparcos}} based cluster 
Hertzsprung-Russel diagram. They derive an age of 625$\pm$50 Myr. 
The Hyades has a deficit 
of low mass stars almost certainly due to dynamical mass loss, where 
the lower mass stars have escaped the cluster
\citep{reid99a,gizis99,dobbie02a,moraux05}.

\citet{bannister07} have recently identified a number of known field 
L and T dwarfs with parallaxes as escaped Hyads and members 
of the Hyades moving group. This identification was made by requiring 
that the proper motions of these dwarfs pointed to the Hyades convergent 
point. These dwarfs also have a ``moving group distance'' that agrees 
with their parallax determined distance.  They found five L and two 
T dwarfs to be members of the Hyades moving group 
(Table \ref{tab_age:HyadesMG_memb}). \citet{zapatero07} measured 
radial velocities for five of the \citet{bannister07} Hyades moving 
group members. These radial velocities, together with the known proper motions, give 
all three components of the velocity. They confirm that these objects 
have velocity components lying in the 2$\sigma$ velocity ellipsoid 
of the Hyades moving group. However, only one, 2MASS J121711.10$-$031113.1
\citep{burgasser99}, 
has space velocities close to those of the Hyades cluster, making it likely to be an escaped cluster member,
 although
velocity dispersion will of course be larger for low mass objects. 
It is 
therefore probably safer to assume the \citet{bannister07} objects, 
despite forming a tight sequence, are members of the moving group
rather than escaped members of the cluster, and so may not be 
exactly coeval. 
2MASS J020529.40$-$115929.6 
\citep[L7;][]{delfosse97,kirkpatrick99} is not included in 
Table \ref{tab_age:HyadesMG_memb} since it is a known binary 
and possibly a triple system \citep{bouy05} whose individual
components are not accurately measured. 
Similarly, the T6 dwarf 2MASS J162414.36$+$002915.8 \citep{strauss99}
is not listed in Table \ref{tab_age:HyadesMG_memb} since if a member, 
it is also probably a binary. In addition, its velocity components are 
not as well defined as those of 2MASS J121711.10$-$031113.1\@.

%
%
\subsection{The Ursa Major moving group}
\label{age:UMa}
The principal members of the Ursa Major moving group are the stars 
that make up the ``plough'', except for $\alpha$ UMa at a $Hipparcos$ distance 
of 38 pc. Group members can be found all over the sky, 
for example Sirius, which may be a member, is only 2.65 pc from 
the Sun (distance from Hipparcos).
Thus the Sun, while not a member, is actually situated inside the 
moving group. The convergent point of the Ursa Major moving group is located at 
$\alpha$=20$^{\rm h}$18.83$^{\rm m}$, $\delta$=$-$34$^{\circ}$25'.8 
\citep{madsen02}.
 The age of the group was determined as 300 Myr by \citet{soderblom93}, although more recently, 
\citet{castellani02} found 
400 Myr, while \citet{king03} have reported an age of 500$\pm$100 Myr 
for the group. 
We use 400 Myr as the age of the moving group.
Using a similar method to that used for the Hyades 
moving group, \citet{bannister07} have found that three L dwarfs and 
one T dwarf with known distances belong to the Ursa Major 
moving group (Table \ref{tab_age:UMa_memb}). There is no evidence 
from the literature that any of these dwarfs are binaries from
high-resolution imaging carried out from space or with ground-based
adaptive optics. One L dwarf, 2MASS J152322.63$+$301456.2 
\citep[also known as Gl\,584C;][]{kirkpatrick00}  has a spectral 
type of L8 and is on the LT transition, or the blueward part of the L 
dwarf sequence.

%
%
\begin{table}
\begin{center}
  \caption{Summary of the one early-L dwarf, two transition objects, 
and one T dwarf, members of the Ursa Major moving group. Magnitudes 
from 2MASS and infrared spectral types are listed for each object 
along with its distance (and its associated error) derived from
parallax measurement.}
  \label{tab_age:UMa_memb}
  \begin{tabular}{@{\hspace{0mm}}l c c c c c@{\hspace{0mm}}}
  \hline
2MASS J\ldots{}  &   $J$  &   $H$  & $K_{s}$ & d$_{p}$       & SpT   \\
  \hline
03454316$+$2540233 & 13.997 & 13.211 & 12.672 & 27.0$\pm$0.4 & L1    \\
14460061$+$0024519 & 15.894 & 14.514 & 13.935 & 22.0$\pm$1.5 & L/T    \\
15232263$+$3014562 & 16.056 & 14.928 & 14.348 & 18.6$\pm$0.4 & L8    \\ 
02431371$-$2453298 & 15.381 & 15.137 & 15.216 & 10.7$\pm$0.4 & T6    \\
\hline
\end{tabular}
\end{center}
\end{table}

%
%

\begin{figure}
   \centering
   \includegraphics[width=\linewidth]{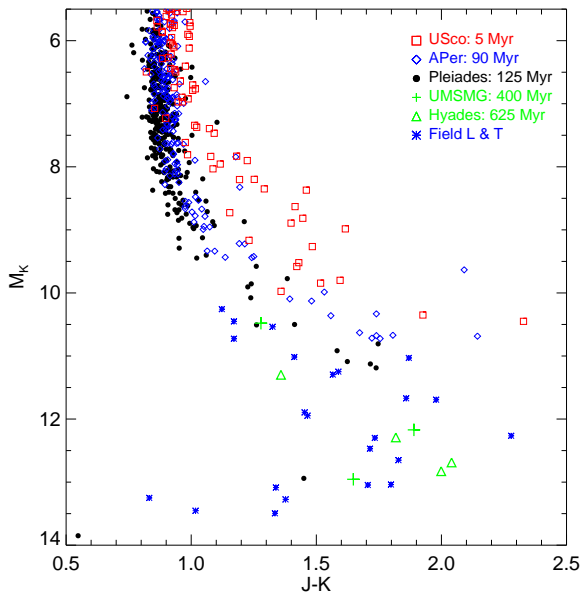}
   \caption{($J-K$,M$_{K}$) colour-magnitude diagram for all the
samples: Upper Sco (squares), \APer{} (diamonds), Pleiades (filled circles),
Ursa Major (crosses), Hyades (triangles), and field dwarfs (star symbols).The photometry is on the MKO system.
}
   \label{fig_age:CMD}
\end{figure}
%

%
%
\subsection{The Pleiades cluster}
\label{age:Pleiades}

The Pleiades is arguably the best studied open cluster in the
northern sky, mainly due to its proximity and richness.
Its mean distance is estimated to be 132$\pm$7 pc. This has been determined using 
various methods, including isochrone fitting 
\citep[126$\pm$6 pc;][]{johnson57}, ground-based parallaxes
\citep*[131$\pm$7 pc;][]{gatewood00} and a recent estimate from 
a detached eclipsing binary member of the Pleiades 
\citep*[139.1$\pm$3.5 pc;][]{southworth05}.
The uncertainty on the distance is roughly half the 
tidal radius of the cluster \citep[13 pc;][]{pinfield00}.
The age of the Pleiades is 125$\pm$8 Myr \citep{stauffer98} as derived 
by the lithium depletion boundary \citep{rebolo92}. The turn-off 
main-sequence age is, however, estimated to be $\sim$80 Myr 
\citep{mermilliod81}, a difference also observed in other open clusters 
\citep{jeffries01} including \APer{} \citep{barrado02a}. The large 
mean proper motion ($\mu \sim$ 50 mas/yr)of the cluster makes the Pleiades an ideal 
place to identify members over a baseline of a few years.

Our sample of Pleiades brown dwarfs include photometric and proper 
motion members extracted from cross-correlations between optical 
and near-infrared data. The full selection procedure is detailed 
in \citet{lodieu07a}. Briefly, we have cross-correlated optical
data from the Issac Newton \citep{pinfield00,dobbie02b} and 
Canada-France-Hawaii \citep{bouvier98,moraux03} surveys with the 
UKIDSS Galactic Cluster Survey (GCS) Data Release 1 available in the 
Pleiades to measure proper motions for about 60 brown dwarfs down to
0.03 M$_{\odot}$ thanks to a five to seven year baseline between the 
optical and infrared observations. Moreover, the stellar sequence 
originates from the work by \citet{adams01a} using the 2MASS database 
\citep{cutri03}. The subsample of low-mass stars and brown dwarf
members fainter than an absolute magnitude M$_{K}$ = 6 mag (MKO) is plotted
as filled dots in Fig.\ \ref{fig_age:CMD}. 

In addition to the above sample, we have added two candidate members
(filled dots in Fig.\ \ref{fig_age:CMD}) with photometry in two 
optical filters from the Canada-France-Hawaii \citep{bouvier98,moraux03} survey 
and in the $JHK$ passbands as well as proper motion 
consistent with the Pleiades measured again over a 5 year baseline \citep{casewell07}.
The first one, PLZJ\,23, is a transition object with $J-K$ (MKO) = 1.45 mag
and a photometric spectral type estimated to be L8--T1.5. The
second object, PLZJ\,93, is a mid-T dwarf with $J-K$  (MKO)= 0.55 mag
\citep{casewell07}.

%
%
\subsection{The $\alpha$ Per cluster}
\label{age:APer}

The $\alpha$ Per cluster has been extensively studied using photometric,
proper motion, and spectroscopic criteria
\citep{stauffer85,stauffer89,prosser92,prosser94}.
The cluster has a distinguishable proper motion of $\mu \sim$ 25 mas/yr
and is 182$\pm$8 pc away from the Sun. This distance represents
a compromise between the {\it{Hipparcos}}
\citep[190.5$^{+7.2}_{-6.7}$ pc;][]{robichon99}
and main-sequence fitting distances
\citep[176.2$\pm$5.0 pc;][]{pinsonneault98}.
The uncertainty in the distance does not affect the absolute magnitudes
by more than $\pm$0.1 mag and is smaller than the tidal radius of
the cluster. The age of the cluster is 90$\pm$10 Myr
from the lithium depletion boundary \citep{stauffer99}, twice the age inferred from the turn-off main-sequence method
\citep[50 Myr;][]{mermilliod81}. \citet*{barrado04} revised the age 
to 85$\pm$10 Myr, the value that we adopt here.

All \APer{} members down to the hydrogen-burning limit are photometric
and proper motion members identified by \citet{deacon04} and confirmed
as such by the GCS data. As for the Pleiades, the lowest mass members 
were extracted from the UKIDSS GCS DR1 over a nine square degree area 
but remain photometric candidates only because we lack proper motions 
due to the small overlap between optical \citep{barrado02a} and infrared 
(GCS DR1) observations. In that respect, this is the most uncertain 
sequence (diamonds in Fig.\ \ref{fig_age:CMD}) among the five
clusters/groups considered here, although we do have photometry in
five passbands ($ZYJHK$). Moreover, the analysis of the control fields 
from the GCS (about three square degrees located roughly four degrees
away from the cluster centre) show that our sequence should suffer from 
a low level of contamination below $K_{s}$ (2MASS)=14.3 mag. 

%
%
\subsection{The Upper Sco association}
\label{age:USco}

Upper Sco is part of the nearest OB association, Scorpius-Centaurus 
\citep{deGeus89} located at a distance of 145 pc from {\it{Hipparcos}}
parallax measurements \citep{deBruijne97,deZeeuw99}. 
The cluster is 5 Myr old with a scatter of less than 2 Myr 
\citep*{preibisch99}. This age estimate is derived from the 
nuclear age \citep[5--6 Myr;][]{deGeus89} and the dynamical age
\citep[4.5 Myr;][]{blaauw91}. Members with masses as low as 0.1 M$_{\odot}$
have been identified by various surveys at multiple wavelengths
\citep{walter94,preibisch98,kunkel99,preibisch01} and an estimate
of the Initial Mass Function \citep{salpeter55} over the
2.0--0.1 M$_{\odot}$ mass range was provided by \citet{preibisch02}.
Several additional optical surveys complemented by near-infrared
photometry have been carried out in the region to find lower mass
stellar members and brown dwarfs \citep{ardila00,martin04,slesnick06}.

We have extended the cluster mass function down to 0.01 M$_{\odot}$
and extracted twenty new brown dwarfs below 0.030 M$_{\odot}$ over
6.5 square degrees surveyed during the science verification phase
of the UKIDSS GCS \citep{lodieu07b}. All photometric candidates
extracted from that survey are displayed as open squares in
Fig.\ \ref{fig_age:CMD}. We have recently confirmed 18 out of 20
photometric candidates (i.e. success rate of 90\%) fainter than 
$K$(MKO) $\sim$ 13.7 mag
(or M$_{K}$(MKO) $\sim$ 7.9 mag) as spectroscopic members based on the their 
spectral shape and the presence of weak gravity-sensitive features 
\citep{lodieu07c}. This limit in magnitude corresponds to effective 
temperatures cooler than 2700\,K and masses below 0.030 M$_{\odot}$
according to theoretical models \citep{chabrier00}. All spectroscopic
members have spectral types later than M8, the majority being
L dwarfs.

%
%
\section{The determination of the age}
\label{age:determination}

In this section we describe the method to derive the age of L dwarfs
using their $J-K$ colours and absolute $K$ magnitudes. Initially we 
consider only those L dwarfs on the redward sequence i.e.\ dwarfs with 
spectral types from L0 up to approximately L8. We discuss the late-L 
and T dwarfs that lie on the L to T blueward sequence in 
Sect.\ \ref{age:transition}. Figure \ref{fig_age:cmd_zoom} shows an 
expanded simplified version of Fig.\ \ref{fig_age:CMD}.
Figure \ref{fig_age:cmd_zoom} shows that the single L dwarfs for each cluster fit a well
defined sequence or isochrone. This is expected theoretically \citep{chabrier00,burrows01,burrows06}. 
For each cluster/isochrone there is a modest decrease ($\sim$ a factor 2) in mass, but a smaller 
decrease ($\sim$ 25 per cent) in radius for the L dwarfs. $J-K$ increasing from 1.0 to 2.0 corresponds 
to a drop in T$_{\rm eff}$ from $\sim$2100 to $\sim$1700 K, which is approximately the same for all of the 
clusters.
The separation of the cluster isochrones in M$_{K}$ is caused by the different radii/surface areas at a given $J-K$/T$_{\rm eff}$.
%
%
\subsection{Restrictions on the method}
\label{age:restrictions}

The first step is to remove known binaries or objects that clearly 
lie above the single star sequence in each cluster/group 
i.e.\ photometric multiple system candidates. Potential
multiple systems in the Pleiades are listed in Appendix F
in \citet{lodieu07a}. For the Hyades, we remove one known binary,
2MASS J0205293$-$115930 (L5.5) identified by \citet{bouy05}.
For $\alpha$ Per, we have excluded candidates lying clearly
above the single star sequence in a similar way as in the Pleiades. 
Finally, we have excluded the reddest dwarf in the Upper Sco sample 
($J-K \sim$ 2.3 mag and M$_{K} \sim$ 11 mag (MKO magnitudes)).

The main restrictions on the method presented in 
Sect.\ \ref{age:method} are:
\begin{enumerate}
\item The method is based on the MKO photometric system: corrections
from the 2MASS into the MKO system are made using the 
transforms of \citet{stephens04}
\item A parallax or other distance determination is needed
\item The object must not be a binary (see above). If it is, 
then the method can only be used if some further information is 
available to determine the photometry of each individual component
\item Its metallicity must be approximately solar 
(see Sect.\ \ref{age:metal})
\item The method is currently valid only for objects with spectral 
type earlier than L8, i.e.\ objects not on the L/T transition 
sequence, nor on the turning point to the L/T transition sequence
(Sect.\ \ref{age:transition})
\item As can be seen from figure \ref{fig_age:AGE} the method will become innacurate for ages $\geq$ 0.7 Gyr
since the isochrones for older  L dwarfs become progressively closer. This is expected theoretically, see \citet{chabrier00} for more details.
\end{enumerate}

%
%
\subsection{Method}
\label{age:method}

\begin{figure}
   \centering
 \scalebox{0.6}{\includegraphics[width=\linewidth,angle=270]{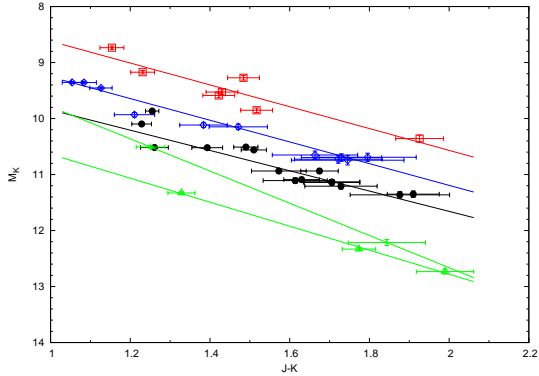}}
   \caption{The ($J-K$,M$_{K}$) diagram for the
members of all five clusters/groups under study, including Upper Sco 
in red (boxes), \APer{} in blue (diamonds), the Pleiades 
in black (filled circles), Ursa Major (crosses) 
and the Hyades in green (triangles). The photometry is on the MKO system.
}
   \label{fig_age:cmd_zoom}
\end{figure}

First of all, we fit each dwarf sequence with a straight line of the 
form M$_{K}$ = A$+$B$\times$($J-K$) (on the MKO system) as shown in 
Fig.\ \ref{fig_age:cmd_zoom}. At the blue end of the sequence,
we terminate the lines where they begin to curve up towards the M dwarf 
sequence. Values of the gradient B for each group are presented in 
Table \ref{tab_age:age_table} along with values of M$_{K}$ at 
$J-K$ = 1.5 mag (MKO). We can see that the fit to the Ursa Major data points is clearly
discrepant with the other groups. Its gradient of 2.88 is steeper 
than the mean values observed for the other clusters but is based
on only two points. 

\begin{figure}
   \centering
 \scalebox{0.3}{\includegraphics[angle=270]{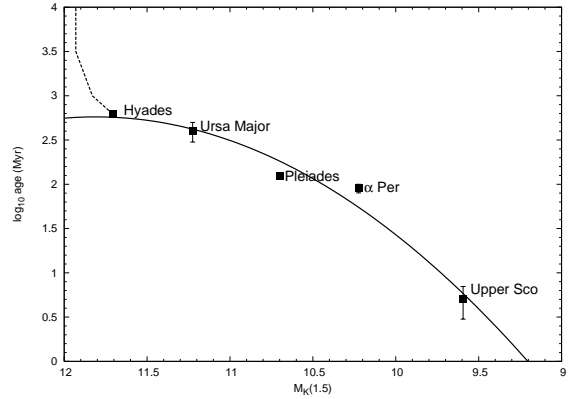}}
   \caption{Age vs M$_{K}$ for $J-K$=1.5 mag.
The clusters are marked by filled squares with error bars corresponding
to uncertainties quoted in Table \ref{tab_age:age_table}. The dotted line
represents the predicted deviation from the quadratic fit (solid line). 
The data for these points are found in Table \ref{tab_age:age_table}.The photometry is on the MKO system. 
}
   \label{fig_age:AGE}
\end{figure}

It would probably be better to use a gradient of 
1.98 around 400 Myr, the average of the ages of the Pleiades and Hyades.
However, the value of M$_{K}$ at $J-K$ = 1.5 mag (MKO) in Fig.\ \ref{fig_age:AGE}
seems reasonable.

We suggest using the following method for determining the age 
of a dwarf with a known absolute magnitude M$_{K}$ and $J-K$ colour. 
Firstly, plot the dwarf on Fig.\ \ref{fig_age:cmd_zoom}, and determine 
which cluster/group it is nearest.
Then use Table \ref{tab_age:age_table} to find the appropriate value 
of B for that cluster, hence the value of M$_{K}$ can then be 
determined corresponding to $J-K$ = 1.5 mag (MKO), 
i.e.\ M$_{K}$(1.5) = M$_{K}-$B($J-K-$1.5). The age in Myr for this 
value of M$_{K}$ at $J-K$ = 1.5 mag can then 
be read off the fit to the data in 
Fig.\ \ref{fig_age:AGE}, given by a quadratic expression shown in equation \ref{1}, 

\begin{eqnarray}
\log(age)& = &-53.9753+9.60944\times M_{K}(1.5)\nonumber\\
&& -0.406885\times M_{K}(1.5)^{2}. 
\label{1}
\end{eqnarray}

As can be seen from Fig.\ \ref{fig_age:AGE} it is possible to
fit a smooth curve through the points that only just misses the
Pleiades and \APer{}.
There is no reason to believe that a smooth curve is the correct
fit, but we will adopt this parabolic fit until more information
is available. Indeed Fig.\ 9 of \citet{burrows01} which plots
dwarf masses versus age and T$_{\rm eff}$ has some kinks around
30 and 100 Myr due to deuterium burning which may explain the anomalous Pleiades
and \APer{} points.
A comparison with theoretical models is probably too premature.
The Lyon group is in the process of revising the DUSTY models
\citep[F.\ Allard, personal comm.;][]{chabrier00} and the Arizona 
group \citep{burrows06} do not calculate isochrones which makes it 
difficult to compare their L dwarf models with observations.

The polynomial (equation \ref{1}) describing the smooth curve in Fig.\ \ref{fig_age:AGE} 
cannot be used for ages greater than the age of the Hyades. For ages $\sim$1 Gyr
to 10 Gyr, gravity will only change by very small amounts,
hence one would expect the surface brightness to be constant
for a given effective temperature (T$_{\rm eff}$) and M$_{K}$ would 
simply scale as 5$\times \log$(radius). Using the DUSTY models from 
\citet{chabrier00} to find radii at T$_{\rm eff} \sim$1800\,K 
(corresponding $J-K$ = 1.5 mag) we can then calculate M$_{K}$ 
relative to the Hyades. This is shown as the dashed line in 
Fig.\ \ref{fig_age:AGE}.
The associated M$_{K}$ values at 
$J-K$ = 1.5 mag are given in Table \ref{tab_age:age_table}.
Clearly this abrupt change of slope at the Hyades is
unsatisfactory. It is tempting to suggest that the age of the
Hyades is older than our adopted value of 625 Myr. This would
make for a smoother gradient change.

\begin{figure}
   \centering
 \scalebox{0.6}{\includegraphics[width=\linewidth,angle=270]{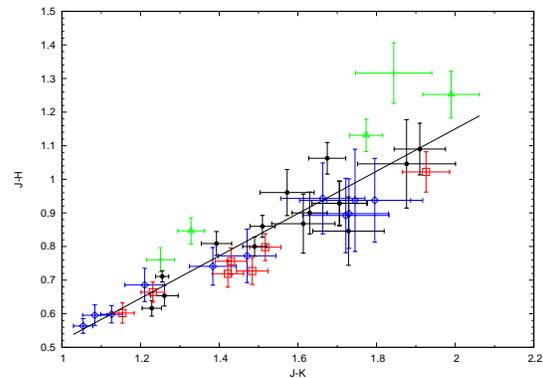}}
   \caption{The ($J-K$, $J-H$) colour-colour diagram for all
members belonging to the five clusters/groups under study, including
Upper Sco 
in red (boxes), \APer{} in blue (diamonds), the Pleiades 
in black (filled circles), Ursa Major (crosses) 
and the Hyades in green (triangles). The photometry is on the MKO system.}
   \label{fig_age:jhhk}
\end{figure}

Figure \ref{fig_age:jhhk} shows the ($J-K$,$J-H$) colour-colour 
diagram for all the single stars used in determining the cluster fits on the MKO system. 
The line can be described by a linear regression like 
$J-H$ = a$+$b$\times$($J-K$), where a = $-$0.11$\pm$0.07 and 
b = 0.63$\pm$0.04. This fit can be used to determine the $K$
magnitude if for some reason only $J$ and $H$ photometry are available. 
One can then proceed as explained above to determine the age of the object.

%
%

%
%
\begin{table}
\begin{center}
  \caption{Name, age (in Myr), gradient B of the fit to M$_{K}$ as
a function of $J-K$ (MKO) and the value M$_{K}$ for $J-K$ = 1.5 mag  (MKO) obtained
from the straight line fits for the five clusters/groups studied
(Fig.\ \ref{fig_age:cmd_zoom}).
$^{a}$ A gradient of 1.98 would be more adequate for Ursa Major
(see discussion in Sect.\ \ref{age:method}).}
  \label{tab_age:age_table}
  \begin{tabular}{@{\hspace{0mm}}l c l c @{\hspace{0mm}}}
  \hline
Name\ldots{}  &   Age  &   Gradient  & M$_{K}$(1.5)  \\
  \hline
Upper Sco  &5$\pm$2     & 1.95$\pm$0.34       &  9.59$\pm$0.03 \\
\APer      &85$\pm$10   & 1.93$\pm$0.12       & 10.22$\pm$0.03 \\
Pleiades   &125$\pm$8   & 1.81$\pm$0.19       & 10.70$\pm$0.04 \\
Ursa Major &400$\pm$100 & 2.88$\pm$0.00$^{a}$ & 11.22$\pm$0.00 \\
Hyades     &625$\pm$50  & 2.14$\pm$0.09       & 11.70$\pm$0.03 \\
-          &   1000     &    -                & 11.83           \\
-          &   5000     &    -                & 11.93           \\
-          &  10000     &    -                & 11.93           \\
\hline
\end{tabular}
\end{center}
\end{table}
%

%
%

%
%
%

%
%
\subsection{Metallicity}
\label{age:metal}

It is clear that for a given spectral type, estimated from the spectrum
over the 0.6 to 0.9 micron wavelength range, the $J-K$ colour is much 
bluer for metal poor dwarfs \citep{burgasser07d}. There is a similar effect seen in M dwarfs where the $J-K$ colour is bluer 
by less than 0.7 mag \citep{gizis97}. This result is confirmed
theoretically i.e.\ $J-K$ gets bluer with decreasing metallicity at
constant effective temperature \citep*{burrows06}. All the clusters
discussed here have approximately solar metallicity, implying that
our age determination can only be valid for solar metallicity dwarfs.
Only a few metal poor dwarfs of L spectral type have been reported
to date \citep{burgasser07b} so extending our method to low metallicity
is currently impossible. To understand the effect of metallicity 
theoretical isochrones ($J-K$,M$_{K}$) are needed for differing 
metallicities at cool temperatures and these are not yet available.

%
%
\subsection{The L to T transition}
\label{age:transition}

\begin{figure}
   \centering
 \scalebox{0.6}{\includegraphics[width=\linewidth,angle=270]{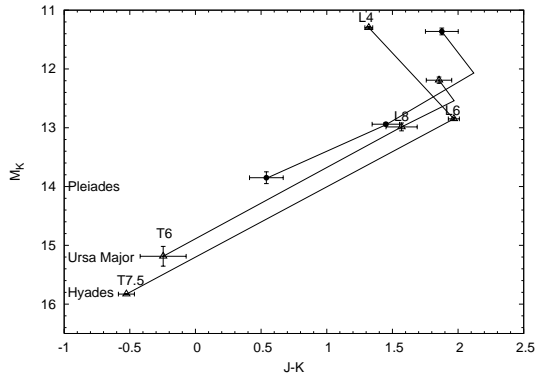}}
   \caption{The ($J-K$,M$_{K}$) CMD for the 
three clusters/groups studied with known L/T transition members. 
The Hyades and Ursa Major are plotted with triangles and the Pleiades objects are plotted with filled circles.
The objects 
with known spectral types are marked (see Tables 
\ref{tab_age:HyadesMG_memb} \& \ref{tab_age:UMa_memb}).
No spectra have been observed for the Pleiades objects, and so no spectral types are available for them.The photometry is on the MKO system.
}
   \label{fig_age:cmd_transition}
\end{figure}

The L dwarf sequence presents two parts.  Up to a spectral type of
$\approx$L8 the $J-K$ colour keeps increasing. It then turns over for
cooler objects, yielding a decrease in the $J-K$ colour due to
the loss of condensate grains from the photosphere and the strengthening methane absorption. Figure \ref{fig_age:cmd_transition} 
shows the change and the blueward transition region for the 
Pleiades (filled squares), Ursa Major (filled triangles) and 
the Hyades (filled diamonds). The Pleiades dwarfs are 
from \citet{casewell07} and the dwarfs for Ursa Major and the 
Hyades come from \citet{bannister07}. As far as we are aware, 
no data exists for the LT transition region for \APer{}
and Upper Scorpius. We have tentatively drawn sequences for the 
three clusters as displayed in Fig.\ \ref{fig_age:cmd_transition}.
As expected, the Pleiades lies above Ursa Major, which is 
itself above the Hyades, suggesting that the trend observed for
L dwarfs extends all the way down to T dwarfs. However we feel that
it would be too premature to claim an age relation for these 
transition L  and T dwarfs, as data are currently limited to a small
number of objects. Clearly, more members are required at cooler
temperatures for all of the clusters to extend this method around 
the maximum $J-K$ point (LT transition) and onto the T sequence 
(Fig.\ \ref{fig_age:cmd_transition}).

%
%

%
%
\section{Application of the method}
\label{age:application}
In this section we apply our method to a number of interesting 
objects reported in the literature, assuming that they have 
solar-like metallicity and that they are single. 

\subsection{The L dwarf binary 2MASS J074642.5$+$200032.1}

2MASS J0746425$+$2000321 is a resolved L0$+$L1.5 binary.
\citet{bouy04} have measured the orbit of this binary
and determine a total mass of 0.146 M$_{\odot}$. They
estimate the individual masses as 0.085 and 0.066 M$_{\odot}$
with uncertainties smaller than 0.01 M$_{\odot}$.
They estimate the age of this system to be 150-500 Myr based on
its position in a Hertzsprung-Russell diagram using the DUSTY \citep{chabrier00} isochrones. \citet{gizis06}
argue that it is an older system with an age of greater than 1 Gyr. They argue that the DUSTY models do not accurately model the complex molecular absorption and grains at this point, and so claim a better method is to compare the colours of  2MASS J074642.5$+$200032.1 with those of field objects, and so determine the system is older.

The system has a measured parallax of 81.9$\pm$0.3 mas/yr, 
corresponding to a distance of 12.2$\pm$0.1 pc and implying
absolute $K$ magnitudes of 10.61$\pm$0.09 and 11.03$\pm$0.17
for the primary and the secondary, respectively.
Assuming $J-K$ colours of 1.12 and 1.3 mag (2MASSS), our method (using the transforms given in \citet{stephens04}) gives ages
of 574 Myr and 575 Myr for the primary and the secondary, respectively.
These results seems to favour the younger ages from previous 
studies. This system is important since with a known age and 
distance, it could be used to directly test theoretical models.

\subsection{The 2MASS J120733.4$-$393254AB system}

2MASS J120733.4$-$393254b (hereafter 2M1207b) is thought to be a
planetary-mass \citep[3--10 M$_{\rm Jup}$;][]{chauvin04,chauvin05}
companion to 2M1207A \citep{gizis02} which belongs to the TW Hydrae
association whose age is estimated to $\sim$8 Myr
\citep{song03,zuckerman04}. Recently, three independent trigonometric
parallaxes yielding consistent values were reported, implying a
distance of 52.4$\pm$1.1 pc \citep{gizis07,biller07,ducourant07}.
Hence, the absolute $K$ magnitude of 2M1207b is M$_{Ks}$ = 13.33
based on the latest near-infrared photometry ($J$ = 20.0$\pm$0.2 mag,
$Ks$ = 16.93$\pm$0.11 mag in the 2MASS system) obtained by
\citet{mohanty07}.

Because 2M1207b has now an accurate age and distance, it would
be a very interesting object to test our method.
Unfortunately its red $J-K$ colour ($J-K$ = 3.07 mag) possibly
due to the presence of a disk seen edge-on \citep{mohanty07}
prevents us from comparing 2M1207b with our samples of L dwarfs
whose $J-K$ colour are bluer than 2.0 mag. We can only put
an upper limit on its age by extending the cluster sequences
in Fig.\ \ref{fig_age:CMD} and argue that it is younger than
\APer{} \citep[85 Myr;][]{barrado04}.

\subsection{The HD\,203030AB system}

HD\,203030B is a late-type (L7.5$\pm$0.5) brown dwarf companion
\citep{metchev06} to HD\,203030A, a young G8 star at 40.8$\pm$1.8 pc 
\citep{jaschek78,perryman97}. The age of the primary is fairly 
well constrained, around 130--400 Myr, with a mean value of 250 Myr based
on photometry, stellar rotation, chromospheric activity, lithium and 
space motion. The secondary has M$_{Ks}$ = 13.15$\pm$0.14 and 
$J-K$ = 1.92 (2MASS)\citep{metchev06}, (M$_{K}$=13.13,$J-K$=1.72, MKO). 
Our method then yields an age of 401 Myr in agreement with the age range of the 
primary star.

%
%
\begin{table*}
\begin{center}
  \caption{Name, absolute $K$ magnitude (M$_{K}$),
$J-K$ colour (MKO), and value of M$_{K}$ at $J-K$ = 1.5,
and estimated ages assuming distances of 300 pc,
360 pc and 440 pc for four spectroscopic L dwarfs
member of the $\sigma$ Ori cluster.
}
  \label{tab_age:sigmaori}
  \begin{tabular}{@{\hspace{0mm}}l c c c @{\hspace{2mm}}c @{\hspace{2mm}}c c c @{\hspace{2mm}}c @{\hspace{2mm}}c c @{\hspace{2mm}}c @{\hspace{2mm}}c @{\hspace{0mm}}}
  \hline
Name &  RA    & dec   & \multicolumn{3}{c}{M$_{K}$} & $J-K$ &  \multicolumn{3}{c}{M$_{K}$($J-K$ = 1.5)}  & \multicolumn{3}{c}{Age (Myr)} \\
  \hline
     &  J2000 & J2000 &  300 pc   &  360 pc  & 430 pc  &  &   300 pc  &  360 pc  & 430 pc   & 300 pc & 360 pc  &  430 pc \\
\hline
SOri\,52&  05 40 09.2& $-$02 26 32.0 &   8.884 &   8.488  &  8.103  &  1.090 &   9.683  &  9.287 &   8.902 &   8.5 &   1.5 &   0.2\\
SOri\,56&  05 39 00.9& $-$02 21 42.0 &   9.470&    9.074  &  8.689  &  1.490 &   9.490 &   9.094 &   8.708 &   3.7  &  0.6 &   0.1\\
SOri\,58&  05 39 03.6& $-$02 25 36.0 &   9.658 &   9.262  &  8.877  &  1.280 &  10.087 &   9.691 &   9.306 &  35.9 &   8.6 &   1.6\\
SOri\,60&  05 39 37.5& $-$02 30 42.0 &  10.197 &   9.801  &  9.416 &   1.350&   10.489 &  10.093 &   9.708 & 113.1 &  36.7 &   9.3\\
SOri\,62&  05 39 42.1& $-$02 30 31.0 &  10.474 &  10.078  &  9.693 &   1.590 &  10.299 &   9.903 &   9.517 &  68.2 &  19.2  &  4.2\\
SOri\,65&  05 38 26.1& $-$02 23 05.0 &  11.434 &  11.038  & 10.653 &   1.080 &  12.253 &  11.857 &  11.472 & $>$580 & 576.1&  519.2\\
SOri\,66&  05 37 24.7& $-$02 31 52.0 &  10.991 &  10.595  & 10.210 &   1.600 &  10.796 &  10.400 &  10.015 & 220.9 &  90.0 &  28.4\\
SOri\,67&  05 38 12.6& $-$02 21 38.0 &  11.234 &  10.838  & 10.453 &   1.300 &  11.624 &  11.228 &  10.843 & 559.2 & 421.0 & 241.1\\
SOri\,68&  05 38 39.1& $-$02 28 05.0 &  11.024 &  10.628  & 10.243 &   1.770  & 10.498 &  10.102 &   9.717 & 115.4 &  37.6 &   9.6\\
\hline
\end{tabular}
\end{center}
\end{table*}

%
%
\subsection{The $\sigma$ Orionis dwarfs}

A large sample of brown dwarfs and planetary-mass objects 
\citep{zapatero00} have been found in the $\sigma$ Orionis cluster 
\citep[age = 3--7 Myr; d = 360$^{+70}_{-60}$ pc;][]{zapatero02,brown94}.
Several of them have been confirmed as spectroscopic members of L
spectral type based on their H$\alpha$ emission and gravity features
at optical wavelengths \citep{zapatero00,barrado01}. The sample 
is divided into two groups: four sources whose $J-K$ colour and 
$K$ magnitudes (SOri56, 58, 60, and 66; Table \ref{tab_age:sigmaori})
are   available from the UKIDSS GCS Data Release 2 \citep{warren07a}
and the remaining ones (Table \ref{tab_age:sigmaori}) whose 
near-infrared photometry is available either from \citet{zapatero00} 
or \citet{martin01a}.

The M$_{K}$ for each of these objects has been calculated for
$J-K$ = 1.5 mag using a slope of 1.95 from Fig.\ \ref{fig_age:cmd_zoom}
i.e.\ the slope for Upper Sco, the cluster closest to $\sigma$ Ori
in terms of age.
Table \ref{tab_age:sigmaori} shows a wide range of ages for these dwarfs.
Our method confirms that using distances of 300 to 430 pc, some, but not all of these objects are young.

%
%
\section{Conclusions and outlook}
\label{Pleiades:conclusions}

We have presented a new method to infer the age of young (age $\leq$0.7 Gyr) L dwarfs from their
infrared photometry alone, assuming that their distance is known 
and they are single objects of solar like metallicity. 
\citet{eggen96} finds that for a sample of nearby lower main sequence stars
53 percent have ages of less than 1 Gyr and we would expect a similar result for L dwarfs.
Thus the technique will be useful for many L dwarfs. 

Our method can only be as accurate as the calibrating clusters used
and it is possible that differences in metallicity between the clusters
could have a significant effect on the determination of the age.
It would be very helpful to have some theoretical guidance on how the
metallicity affects the isochrones which would involve running
the structure and atmospheric models for different metallicities.
At the very least the method is capable of saying whether
a given dwarf is older or younger than the calibrating
clusters with an accuracy of $\pm$0.2 in $\log$(age).
However, relative ages can be measured more precisely. 
With a dynamic range $\sim$2.5 magnitudes in absolute magnitude,
it has the potential, with further observations and theoretical input, 
to become a very accurate and powerful method. Hence,
we anticipate that this method will evolve and improve with the
discovery of a larger number of L dwarfs with a wider range of ages
as well as with updated theoretical models.

%
%
\section*{Acknowledgements}

NL, SLC and PDD are postdoctoral research associates that were funded 
for part of this work by PPARC and STFC.
This research has made use of the Simbad database of NASA's
Astrophysics Data System Bibliographic Services (ADS).
Research has benefited from the M, L, and T dwarf compendium 
housed at DwarfArchives.org and maintained by Chris Gelino, 
Davy Kirkpatrick, and Adam Burgasser.

%
%

\bibliographystyle{mn2e}
%
\bibliography{mnemonic,bibliography_extra,nic_refs,test}
\label{lastpage}

\end{document}